\documentclass[a4paper,11pt]{article}

\usepackage{jheppub} 
\usepackage{lmodern}
\usepackage{tikz}
\usetikzlibrary{decorations.pathmorphing}
\tikzset{snake it/.style={decorate, decoration=snake}}

\usepackage{slashed}
\usepackage[]{mdframed}
\usepackage[T1]{fontenc} 

\newcommand{\be}{\begin{equation}}
\newcommand{\bea}{\begin{eqnarray}}

\newcommand{\ee}{\end{equation}}
\newcommand{\eea}{\end{eqnarray}}

\title{\boldmath 
Complexity and Operator Growth in Holographic 6d SCFTs
}
\author[a,b]{Ali Fatemiabhari }
\author[a]{ Carlos Nunez}
\author[a]{and Ricardo T. Santamaria}

\affiliation[a]{Centre for Quantum Fields and Gravity, Department of Physics, Swansea University, Swansea SA2 8PP, United Kingdom}
\affiliation[b]{Institute for Theoretical and Mathematical Physics, Lomonosov Moscow State University, 119991 Moscow, Russia}

\abstract{We study Krylov (spread) complexity in strongly coupled six–dimensional 
${\cal N}=(1,0)$
 superconformal field theories with holographic duals in massive type IIA supergravity. Extending recent holographic proposals relating Krylov complexity growth to the proper momentum of an infalling particle, we analyse the dynamics of massive geodesic probes in these geometries.
In our setup, the bulk particle is allowed to move along three directions: the radial AdS coordinate, the internal 
$S^2$ associated with the 
$SU(2)_R$
  symmetry, and the coordinate parametrising the quiver. In the dual field theory these motions encode, respectively, operator growth, the presence of R-symmetry charges, and spreading across different nodes of the quiver. 
We analyse the geodesic motion both analytically and numerically for representative quiver configurations. The motion along the quiver direction is typically damped and localised at early times, while the late-time behaviour is dominated by the radial AdS motion. As a consequence, the generalised proper momentum grows linearly at late times, consistent with expectations for Krylov complexity in conformal theories. The inclusion of angular momentum ($SU(2)_R$ charge) introduces additional constraints on the allowed motion and modifies the early-time dynamics while leaving the asymptotic behaviour unchanged.
These results provide a first exploration of Krylov complexity in higher-dimensional holographic conformal theories and reveal how operator growth can probe both internal symmetries and quiver structure in strongly coupled  conformal field theories.}

\begin{document} 
\maketitle
\flushbottom
\section{Introduction and General Idea}
Understanding how quantum information spreads in strongly interacting
quantum systems has become a central theme connecting quantum field
theory, quantum information and gravitational physics.
In recent years, ideas from quantum complexity have proven particularly
useful in characterising the growth of operators, thermalisation and
chaotic dynamics in many-body systems. For recent review articles, see \cite{Baiguera:2025dkc, Rabinovici:2025otw, Nandy:2024evd}.

In holography \cite{Maldacena:1997re}, where strongly coupled dynamics can be
described geometrically, these ideas have led to a fruitful interplay
between gravitational physics and quantum information.
In the context of gauge/gravity duality, several proposals relate the
complexity of quantum states in the boundary field theory to geometric
quantities in the dual gravitational description.

These ideas have motivated an extensive literature studying complexity
in holography and quantum field theory.
Recently, Krylov complexity has emerged as a particularly natural
measure of operator growth in quantum many-body systems
\cite{Parker:2018yvk}.
In this framework one constructs the Krylov basis generated by repeated
action of the Hamiltonian on a given state or operator.
The resulting Lanczos coefficients determine an effective
one-dimensional chain describing the evolution of the system, and the
spread of the wavefunction along this chain defines the Krylov
complexity \cite{Parker:2018yvk, Rabinovici:2023yex}.
These ideas have been explored in a variety of systems including
conformal field theories and holographic models
\cite{Caputa:2021sib, Caputa:2024xkp,Avdoshkin:2022xuw,Barbon:2019yrr, Jeong:2026iac}.

A particularly interesting holographic proposal was put
forward in \cite{Caputa:2024sux}, where the growth rate of Krylov complexity
in holographic two-dimensional conformal field theories was related to
the proper radial momentum of a falling particle in the dual bulk
geometry.
In this picture, operator growth in the boundary theory is mapped to the
motion of particles following geodesics in the gravitational
background.
This proposal provides an appealing geometric description of operator
spreading and raises the question of how such ideas extend to more
general holographic systems, particularly to higher dimensional
strongly coupled quantum field theories.

Six-dimensional $\mathcal{N}=(1,0)$ superconformal field theories
provide a particularly interesting setting to explore these
questions.
These theories arise in string theory through Hanany-Witten \cite{Hanany:1996ie} type brane
constructions involving NS5, D6 and D8 branes
\cite{Hanany:1997gh, Gaiotto:2014lca}.
Although many of these theories do not admit a conventional Lagrangian
description, they possess well understood holographic duals in massive
type IIA supergravity in terms of $AdS_7$ backgrounds
\cite{Cremonesi:2015bld,Apruzzi:2013yva, Apruzzi:2015wna,Passias:2015gya}.
The geometry of these solutions is characterised by a function
$\alpha(\eta)$ whose derivatives encode the rank function of a linear
quiver describing the tensor branch of the theory
\cite{Cremonesi:2015bld}.
These systems therefore provide a rich framework in which strongly
coupled dynamics, brane physics and holography can be studied
simultaneously. The entanglement entropy in field theories dual to these backgrounds and their deformations are studied in \cite{Chatzis:2025wfv}.

In this paper we investigate Krylov complexity in the context of these
six-dimensional holographic theories.
Extending the proposal of \cite{Caputa:2024sux,Fatemiabhari:2025poq}, we model the spreading of
operators in the SCFT by the motion of a massive particle following a
timelike geodesic in the dual $AdS_7$ background.
In this construction, motion along the internal $S^2$ directions
encodes the $SU(2)_R$ charge of the operator, which is analogous to symmetry resolved Krylov complexity \cite{Caputa:2025mii, Caputa:2025ozd}. Similarly, motion along the
coordinate $\eta$ corresponds to spreading across different nodes of
the quiver \cite{Fatemiabhari:2025poq}.
We derive the corresponding equations of motion and analyse them both
analytically and numerically for two representative quiver
configurations.
Our results show that the motion along the quiver direction is typically
damped and largely confined to early times, while at late times the
dynamics is dominated by the radial motion in $AdS$.
Consequently, the generalised proper momentum, which we interpret as
the rate of holographic complexity growth, approaches a linear behaviour
at late times, consistent with expectations for conformal theories.

A word on the dual field-theory interpretation of the geodesic
initial conditions is in order.  In the holographic dictionary, the
massive point particle in the bulk corresponds to a local operator
$\mathcal{O}$ in the 6d SCFT with conformal dimension $\Delta\propto
mL_{\mathrm{AdS}}$ (large $\Delta$ justifying the point-particle
approximation).  The $\mathrm{SU}(2)_R$ charge $J$ corresponds to the
$R$-charge of $\mathcal{O}$, encoded in its transformation properties
under the $S^2$ isometry.  The initial position $\eta_0$ in the quiver
direction localises the operator at a specific gauge node of the quiver:
operators built from fields at node $k$ correspond to $\eta_0\approx k$
in the continuum limit.  More precisely, bifundamental fields connecting
nodes $k$ and $k+1$ are dual to particles that can propagate in the
$\eta$-direction.  The spreading of this operator across the quiver in the
Krylov sense is then mapped to the particle's motion along $\eta(t)$.
For operators of fixed, large $\Delta$ and fixed $J$, the point-particle
action is the appropriate
bulk effective action.  For operators corresponding to extended objects
-- such as instanton strings (D2-branes on $(t,x_1,x_6)$ ending on
NS5-branes) or D-brane excitations -- the appropriate bulk probe
would instead be a Dirac-Born-Infeld-Wess-Zumino action or a string worldsheet action,
respectively.  These would introduce higher-derivative corrections
in $\alpha'$ and could capture complexity growth for operators of
fixed charge but extensive conformal dimension. Whilst some of these issues were discussed in \cite{Nastase:2026lhz}, we leave the
exploration of such extended-probe contributions to Krylov complexity
for future work.

The rest of the paper is organised as follows.
In Section \ref{sec:SCFTs6} we review the holographic description of six-dimensional
$\mathcal{N}=(1,0)$ SCFTs and introduce the two quiver examples that we
will use throughout the paper.
In Section \ref{sec:krylovreview} we review Krylov complexity and extend the holographic
prescription relating complexity growth to generalised proper momentum in the bulk.
We then analyse the resulting geodesic equations for different quiver
configurations and values of the $R$-symmetry charge, we draw some parallels with symmetry resolved Krylov complexity. A careful numeric study is presented in detail.
In Section \ref{conclusionsection} we summarise our results and discuss several directions for
future work.

\section{Super Conformal Theories in Six Dimensions and Holographic Dual}\label{sec:SCFTs6}
Let us start with a short summary of six-dimensional $\mathcal{N}=(1,0)$ conformal field theories and their holographic description.
It is useful to recall the main obstruction for interacting local quantum field theories in dimensions $d>4$. Consider a simple scalar field theory in six dimensions,
\begin{equation}
S=\int d^6x \left[-\tfrac{1}{2}(\partial_\mu \phi)^2 - V(\phi)\right].\nonumber
\end{equation}
A real scalar has classical mass-dimension $[\phi]=m^2$, a mass term reads $V=\tfrac{m^2}{2}\phi^2$. A marginal-looking cubic $V=g\phi^3$ is problematic as it lacks a bounded below potential for $\phi<0$, while a stable $\lambda\phi^4$ interaction is irrelevant in six dimensions, so the theory requires a UV completion. The Wilsonian intuition (start with a UV CFT and perturb by relevant operators) appears not to straightforwardly supply interacting renormalisable models in $d=6$.

Despite this, string constructions indicate that supersymmetric theories of scalars coupled to gauge fields may possess interacting UV fixed points. For example, a schematic Lagrangian
\begin{equation}
\mathcal{L}\sim -\tfrac{1}{2}(\partial_\mu \phi)^2 -c\,\phi F_{\mu\nu}^2 +\text{fermions},
\label{kakit}
\end{equation}
shows that when $\langle\phi\rangle\to0$ one probes the strong coupling limit of a gauge theory, with $\phi$ playing the role of the inverse gauge coupling. Supersymmetry introduces fermions and potential gauge anomalies; anomaly cancellation is possible when $\phi$ sits in a tensor multiplet \cite{Seiberg:1996qx,Danielsson:1997kt}, with a finely tuned matter content.

Hanany--Witten type brane constructions realise such theories, see \cite{Hanany:1996ie,Hanany:1997gh}. The six-dimensional $\mathcal{N}=(1,0)$ theories preserve eight Poincaré supercharges and have bosonic symmetry $SO(1,5)\times SU(2)_R$. The relevant multiplets are:
\begin{itemize}
  \item \textbf{Tensor multiplet:} $(B_{\mu\nu},\lambda_1,\lambda_2,\phi)$ (self-dual two-form, fermions, real scalar).
  \item \textbf{Vector multiplet:} $(A_\mu,\hat{\lambda}_1,\hat{\lambda}_2)$.
  \item \textbf{Hypermultiplet:} $(\varphi_1,\varphi_2,\psi_1,\psi_2)$.
  \item \textbf{Linear multiplet:} $(\vec{\pi},c,\tilde{\xi})$ (an $SU(2)$ triplet, singlet and a fermion).
\end{itemize}
When the tensor scalar acquires a VEV (the \emph{tensor branch}) the $SU(2)_R$ symmetry is preserved. VEVs of hyper/linear multiplets explore the Higgs branch, breaking $SU(2)_R$. In this paper, we focus on the tensor branch.

To realise these theories from branes one distributes NS5, D6 and D8 branes as in Table~\ref{table:BraneSetup}.
\begin{table}[h!]
\centering
\begin{tabular}{c||c c c c c c |c|c c c}
  & $t$ &  $x_1$ & $x_2$ & $x_3$ & $x_4$ & $x_5$ & $x_6$ & $x_7$ & $x_8$ & $x_9$ \\ [0.5ex] 
 \hline\hline
 NS5 & $\bullet$ & $\bullet$ & $\bullet$ & $\bullet$ & $\bullet$ & $\bullet$ & $\cdot$ & $\cdot$ & $\cdot$ & $\cdot$ \\ 
 \hline\hline
  D6 & $\bullet$ & $\bullet$ & $\bullet$ & $\bullet$ & $\bullet$ & $\bullet$ & $\bullet$ & $\cdot$ & $\cdot$ & $\cdot$ \\
  \hline
  D8 & $\bullet$ & $\bullet$ & $\bullet$ & $\bullet$ & $\bullet$ & $\bullet$ & $\cdot$ & $\bullet$ & $\bullet$ & $\bullet$ 
\end{tabular}
\caption{Brane set-up: all branes extend on Minkowski $\mathbb{R}^{1,5}$. D6's are finite along $x_6$ between NS5's; D8's extend along $x_{7,8,9}$ preserving $SO(3)_R$.}
\label{table:BraneSetup}
\end{table}

{
Important differences with lower-dimension Hanany--Witten systems include: the NS5 and bounded D6 worldvolumes have the same dimension so five-brane dynamics do not decouple and tensor multiplets (provided by the NS branes) appear. These tensor multiplets are needed to cancel a gauge anomaly \cite{Hanany:1997gh}. The brane content (analogously, the stability of the brane set-up) is restricted to satisfy
\begin{equation}
N_{D6,R} + N_{D6,L} + N_{D8}= 2 N_{D6,c},
\label{anomalycancellation}
\end{equation}
where $N_{D6,R/L}$ are sixbranes to the right/left of a 'colour' stack of $N_{D6,c}$ branes. 
D2-branes on $(t,x_1,x_6)$ ending on NS5's represent strings charged under the self-dual $H_3$. These strings have a dual role. They are coupled to the tensor multiplets living on the world-volume of the NS-branes and are thus charged both electrically and magnetically to the tensor multiplet. In addition, they are instantons to the gauge fields living on the D6-branes.

Flowing to the IR, when the tensor-branch separations are taken as $\langle\phi_i-\phi_{i-1}\rangle\neq0$, these instantonic strings become massive and the low-energy theory admits an anomaly-free quiver description. }On the other hand, in the tensionless string limit $\langle\phi_i-\phi_{i-1}\rangle\to0$ the system is proposed to be UV-completed by  a strongly coupled 6d CFT \cite{Seiberg:1996qx}. These fixed points have a holographic description that we discuss now.

\section*{Massive Type IIA duals: AdS$_7$ backgrounds and the quiver map}
We now summarise the massive Type IIA supergravity backgrounds dual to these $\mathcal{N}=(1,0)$ SCFTs. The family of solutions was written in \cite{Apruzzi:2013yva,Apruzzi:2015wna}; we follow the notation of \cite{ct2015}. The backgrounds (metric, NS $B_2$, dilaton $\Phi$, RR fields $F_0,F_2$) possess $SO(2,6)\times SU(2)_R$ isometries and, in {\it Einstein frame}\footnote{This choice of frame is motivated by the fact that we probe this background with particles, which couple to the Einstein frame metric.} take the form
\begin{equation}
\begin{aligned}
& ds^2_{E}=f_1(\eta)\, ds^2_{AdS_7}+f_2(\eta)\, d\eta^2+f_3(\eta)\, d\Omega_2(\theta,\varphi),\\
& B_2=f_4(\eta)\,\mathrm{Vol}(S^2),\qquad F_2=f_5(\eta)\,\mathrm{Vol}(S^2),\qquad e^{\Phi}=f_6(\eta),\\
& ds^{2}(\text{AdS}_{7}) = dr^{2} + e^{-2r}(dx^{2}_{1,5}),~d\Omega_2=d\theta^2+\sin^2\theta d\varphi^2, ~\text{Vol}(S^2)=\sin\theta d\theta\wedge d\varphi \, .
\end{aligned}
\label{backgroundads7xm3}
\end{equation}
The functions $f_i(\eta)$ are expressed in terms of a single function $\alpha(\eta)$ and its derivatives,
\begin{equation}
\begin{gathered}
f_1(\eta)= 8\sqrt{2}\pi e^{-\frac{\Phi}{2}}\sqrt{-\frac{\alpha}{\alpha''}},\qquad f_2(\eta)=\sqrt{2}\pi e^{-\frac{\Phi}{2}}\sqrt{-\frac{\alpha''}{\alpha}},\\
f_3(\eta)=\sqrt{2}\pi e^{-\frac{\Phi}{2}} \sqrt{-\frac{\alpha''}{\alpha}}\left(\frac{\alpha^2}{(\alpha')^2-2\alpha\alpha''}\right),\\
f_4(\eta)=\pi\!\left(-\eta+\frac{\alpha\alpha'}{(\alpha')^2-2\alpha\alpha''}\right),\qquad
f_5(\eta)=\left(\frac{\alpha''}{162\pi^2}+\frac{\pi F_0\alpha\alpha'}{(\alpha')^2-2\alpha\alpha''}\right),\\
f_6(\eta)=2^{5/4}\pi^{5/2}3^4\frac{(-\alpha/\alpha'')^{3/4}}{\sqrt{(\alpha')^2-2\alpha\alpha''}}.
\end{gathered}
\label{functionsf}
\end{equation}
Supersymmetry and the massive IIA equations require $\alpha(\eta)$ to satisfy
\begin{equation}
\alpha'''(\eta) = -162\pi^3 F_0,
\label{alphathird}
\end{equation}
so for piecewise-constant Romans mass $F_0$ one finds $\alpha(\eta)$ as piecewise cubic,
\begin{equation}
\alpha(\eta)=a_0+a_1 \eta +\tfrac{a_2}{2}\eta^2 -\tfrac{162\pi^3 F_0}{6} \eta^3.
\label{alpha}
\end{equation}

Cremonesi and Tomasiello \cite{ct2015} gave a practical recipe mapping balanced linear quivers to an $\alpha(\eta)$ solving eq.\eqref{alphathird}. Introduce the (piecewise-linear) rank function
\begin{equation}
R(\eta)=-\frac{1}{81\pi^2}\alpha''(\eta),
\end{equation}
and consider a linear quiver with $P$ gauge nodes such that
\[
R(\eta)=\begin{cases}
N_1 \eta & 0\le \eta\le1,\\
N_1+(N_2-N_1)(\eta-1) & 1\le \eta\le2,\\
\vdots & \\
N_{P}(P+1-\eta) & P\le \eta\le P+1.
\end{cases}
\]
The ranks of the gauge groups are encoded in $R$ evaluated at integers $\eta=1,\dots,P$, and the balancing condition reads
\begin{equation}
R''(\eta)=\sum_{k=1}^{P} F_k\delta(\eta-k),\qquad F_k=2N_k-N_{k-1}-N_{k+1},
\label{balance}
\end{equation}
so that each $F_k$ gives the number of flavours coupled to node $k$. Given $R(\eta)$ one integrates $-81\pi^2 R(\eta)$ twice to obtain $\alpha(\eta)$, fixing integration constants by continuity of $\alpha,\alpha'$ and imposing $\alpha(0)=\alpha(P)=0$.

\begin{figure}[h!]
\begin{center}
\begin{tikzpicture}[scale=0.85]
\node (1) at (-4,0) [circle,draw,thick,minimum size=1.1cm] {N$_1$};
\node (2) at (-2,0) [circle,draw,thick,minimum size=1.1cm] {N$_2$};
\node (3) at (0,0) {$\dots$};
\node (4) at (2,0) [circle,draw,thick,minimum size=1.1cm] {N$_{P}$};
\draw[thick] (1) -- (2) -- (3) -- (4);
\node (1b) at (-4,-1.8) [rectangle,draw,thick,minimum size=0.9cm] {F$_1$};
\node (2b) at (-2,-1.8) [rectangle,draw,thick,minimum size=0.9cm] {F$_2$};
\node (4b) at (2,-1.8) [rectangle,draw,thick,minimum size=0.9cm] {F$_{P}$};
\draw[thick] (1) -- (1b);
\draw[thick] (2) -- (2b);
\draw[thick] (4) -- (4b);
\end{tikzpicture}
\end{center}
\caption{Linear quiver: balancing implies $F_k=2N_k-N_{k-1}-N_{k+1}$.}
\label{fig:quiver}
\end{figure}
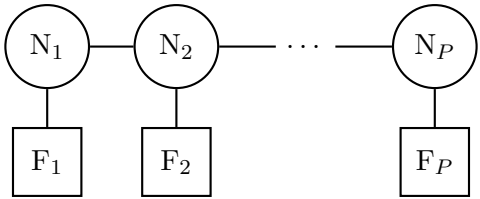

When this $\alpha(\eta)$ is plugged into \eqref{backgroundads7xm3}--\eqref{functionsf} the resulting geometry is the holographic dual to the UV SCFT$_6$ associated to the quiver of Figure~\ref{fig:quiver} (evaluated at the origin of the tensor branch).

Several holographic tests have been performed: Page charges for RR/NS fields map to the quiver and Hanany--Witten brane picture; holographic free energy computations can be compared with field theory expressions \cite{Nunez:2023loo}; see \cite{Nunez:2018ags,Filippas:2019puw} for further observables computed holographically.
\subsection{Two examples}\label{sectwoexamples}
We discuss two examples of the previous formalism. These are used below to evaluate our expression for complexity derived in later sections.
\subsubsection{\underline{\bf Quiver 1}}\label{subsection:quiver1}
For the first case, we chose a quiver that features a flavour group at $\eta=P$, resulting in $\alpha(\eta)$ and a triangular rank function.

$$
R_1(\eta)=-\frac{1}{81\pi^2}\alpha_1''(\eta) = \left\{
 \begin{array}{ll}
N \eta & \quad 0 \leq \eta\leq P, \\
N P (P+1-\eta) & \quad P\leq \eta\leq P+1;
 \end{array}
 \right.
$$
the corresponding quiver is
\begin{center}
	\begin{tikzpicture}
	\node (1) at (-6,0) [circle,draw,thick,minimum size=1.2cm] {N};
	\node (2) at (-4,0) [circle,draw,thick,minimum size=1.2cm] {2N};
	\node (3) at (-2,0) [circle,draw,thick,minimum size=1.2cm] {3N};	
	\node (4) at (0,0) {$\dots$};
	\node (6) at (4,0) [rectangle,draw,thick,minimum size=1.2cm] {(P+1)N};
	\node (5) at (2,0) [circle,draw,thick,minimum size=1.2cm]{PN};
	\draw[thick] (1) -- (2) -- (3) -- (4) -- (5)-- (6);
	\draw[thick] (1,0) -- (1.3,0);
	\draw[thick] (2.7,0) -- (3.3,0);
	\end{tikzpicture}\
\end{center}
The function $\alpha_1(\eta)$ reads,
\begin{eqnarray}
& &\alpha_1(\eta) = \nonumber\\
& &-81\pi^2~N\left\{
 \begin{array}{ll}
\displaystyle a_1 \eta + \frac{1}{6} \eta^3  & \quad 0 \leq \eta\leq P \\   \nonumber\\
\displaystyle(P a_1 +\frac{P^3}{6}) + (a_1+\frac{P^2}{2})(\eta-P) +\frac{P}{2} (\eta-P)^2   -\frac{P}{6}(\eta-P)^3 & \quad P\leq \eta\leq P+1 .      
\label{quiver1}
 \end{array}
 \right.\\
\end{eqnarray}
Where $k:=1,....,P-1$ and $a_1=-\frac{P(P+2)}{6}$. One can check that $\alpha_1(\eta)$ and $\alpha_1'(\eta)$ are continuous and $\alpha_1''(\eta)=-81\pi^2 R_1(\eta)$.

\subsubsection{\underline{\bf Quiver 2}}\label{subsubsection:quiver2}
Here, we choose a quiver with two flavour groups, one at $\eta=1$ and one at $\eta=P$. 
The rank function is, 
$$
R_2(\eta)=-\frac{1}{81\pi^2}\alpha_2''(\eta) = \left\{
 \begin{array}{ll}
N \eta & \quad 0 \leq \eta\leq 1, \\
N & \quad 1\leq \eta\leq P;\\
N (P+1-\eta) & \quad P\leq \eta\leq P+1;
 \end{array}
 \right.
$$
the corresponding quiver is
\begin{center}
	\begin{tikzpicture}
 \node (0) at (-8,0) [rectangle,draw,thick,minimum size=1.2cm] {N};
	\node (1) at (-6,0) [circle,draw,thick,minimum size=1.2cm] {N};
	\node (2) at (-4,0) [circle,draw,thick,minimum size=1.2cm] {N};
	\node (3) at (-2,0) [circle,draw,thick,minimum size=1.2cm] {N};	
	\node (4) at (0,0) {$\dots$};
	\node (6) at (4,0) [rectangle,draw,thick,minimum size=1.2cm] {N};
	\node (5) at (2,0) [circle,draw,thick,minimum size=1.2cm] {N};
	\draw[thick] (0) -- (1) -- (2) -- (3) -- (4) -- (5)-- (6);
	\draw[thick] (1,0) -- (1.3,0);
	\draw[thick] (2.7,0) -- (3.3,0);
	\end{tikzpicture}\
\end{center}
The function $\alpha_2(\eta)$ is
{\small
\begin{eqnarray}
& &-\frac{\alpha_2(\eta) }{81\pi^2 N }= \nonumber\\
& &\left\{
 \begin{array}{ll}\vspace{0.3cm}
\displaystyle a_1 \eta + \frac{\eta^3}{6}   & \quad 0 \leq \eta\leq 1 \\ \vspace{0.3cm}
\displaystyle\left( a_1 +\frac{1}{6}\right) +\left(a_1+\frac{1}{2}\right)(\eta-1)\! + \frac{1}{2}(\eta-1)^2  & \quad 1 \leq \eta\leq 2 \nonumber\\\vspace{0.3cm}
\displaystyle\left( 2a_1 +\frac{7}{6}\right) \!+\!\left(a_1+\frac{3}{2}\right) (\eta-2)\! +\! \frac{1}{2}(\eta-2)^2  & \quad 2 \leq \eta\leq 3 \nonumber\\
\displaystyle\left( 3a_1 +\frac{19}{6}\right) \!\!+\!\!\left(a_1\!+\frac{5}{2}\right)\! (\eta-3)\! +\! \frac{1}{2}\!(\eta-3)^2  & \quad 3 \leq \eta\leq 4 \nonumber\\
... \\
\vspace{0.3cm}
\displaystyle\left(k a_1 +\frac{3k^2 -3k +1}{6}\right) + \left(a_1+\frac{2k-1}{2}\right)(\eta-k) +\frac{1}{2} (\eta-k)^2   , &  k\leq \eta\leq k+1\\  
\displaystyle\left(P a_1 +\frac{3P^2 -3P +1}{6}\right) + \left(a_1+\frac{2P-1}{2}\right)(\eta-P) +\frac{1}{2} (\eta-P)^2   -\frac{1}{6}(\eta-P)^3, &  P\leq \eta\leq P+1,        
 \end{array}
 \right.\\
\end{eqnarray}
}
Here $a_1=-\frac{P}{2}$ and $1\leq k\leq P-1$.
\\
As a summary, we have presented the holographic description of six dimensional ${\cal N}=(1,0)$ SCFTs linear quivers. Papers studying further aspects of these SCFTs are \cite{Apruzzi:2013yva, Apruzzi:2015wna, Apruzzi:2017nck, Apruzzi:2025znw, Nunez:2025gxq, Nunez:2025ppd, Nunez:2025puk,Jokela:2025cyz, Giliberti:2024eii, Conti:2024qgx, Apruzzi:2024ark, Lima:2023ggy, Bergman:2020bvi, Malek:2019ucd}.
We have quoted the function $\alpha(\eta)$ and its second derivative (the rank function), for two special quivers that are generic enough and easy to work with. We use these quivers in the remainder of the paper, when we study specific numerics. 
Let us now focus on the Krylov (spread) complexity.

\section{Krylov (Spread) Complexity}\label{sec:krylovreview}
In this section, we start with a short  introduction to Krylov/spread complexity and the calculation of \cite{Caputa:2024sux}. Then, we generalise this calculation to the six dimensional SCFTs of linear quiver type discussed above. We follow the work of \cite{Fatemiabhari:2025poq}. 
\\
Given a Hamiltonian $H$ and an initial state $|\psi_0\rangle$, the Krylov subspace
$\mathcal{K}$ = $\text{span}\{|\psi_0\rangle,$ $H|\psi_0\rangle,$ $H^2|\psi_0\rangle,\dots\}$
admits an orthonormal {Krylov/Lanczos basis} $\{|n\rangle\}_{n\ge0}$ constructed by the
Lanczos algorithm. Expanding the evolved state as
\begin{equation}
|\psi(t)\rangle=e^{-iHt}|\psi_0\rangle=\sum_{n\ge0}\phi_n(t)\,|n\rangle.
\end{equation}
We have $\sum_n|\phi_n(t)|^2=1$, and the amplitudes obey a discrete Schr\"odinger
equation on an effective 1D chain,
\begin{equation}
i\,\dot\phi_n(t)=a_n\,\phi_n(t)+b_{n+1}\,\phi_{n+1}(t)+b_n\,\phi_{n-1}(t),\qquad b_0:=0,
\end{equation}
where $\{a_n,b_n\}$ are the (real) Lanczos coefficients.
The {spread complexity} (a state-version of Krylov complexity) is the average position
on this chain,
\begin{equation}
\mathcal{C}(t)\equiv \sum_{n\ge0} n\,|\phi_n(t)|^2
\qquad \left(\text{with }\;p_n(t):=|\phi_n(t)|^2\right).
\end{equation}
Operationally one does the following: first, compute the return amplitude $\langle\psi_0|\psi(t)\rangle$. Then,
extract the moments to obtain $\{a_n,b_n\}$. Finally, solve for $\phi_n(t)$, and then evaluate $\mathcal{C}(t)$.

{ The holographic set-up of \cite{Caputa:2024sux} can be described as follows}.
The authors of \cite{Caputa:2024sux} study locally excited states in holographic 2D CFTs,
prepared by inserting a primary operator $\mathcal{O}$ (with a small Euclidean smearing)
into a thermal (TFD) or vacuum background. Conformal symmetry fixes the relevant return
amplitudes in terms of ratios of two-point functions, which in turn leads to analytic
Lanczos data and hence to closed-form expressions for the {rate} $\dot{\mathcal{C}}(t)$
in these 2D CFT examples. The key Physics is that after a characteristic time scale,
$\dot{\mathcal{C}}(t)$ captures the expected rapid growth associated with operator/state
spreading in a holographic theory \cite{Caputa:2024sux}.

{ The proposal of \cite{Caputa:2024sux}--see also \cite{He:2024pox, Fan:2024iop} for elaborations and refinements, is to calculate the time derivative of the  complexity rate as \emph{proper} radial momentum.}
On the gravity side, the same local excitation is semi-classically approximated  by a massive
particle falling along a timelike radial geodesic in an asymptotically AdS$_3$ geometry
(vacuum, global AdS, or BTZ, depending on the CFT state).
The authors of \cite{Caputa:2024sux}  propose to use the radial coordinate $\rho$ measuring
\emph{proper radial distance} (e.g.\ to the horizon in BTZ, or to the Poincar\'e ``horizon''
in vacuum AdS). Writing the relevant 2D $(t,\rho)$ part of the metric as
\begin{equation}
ds^2_{(2)}=-f(\rho)\,dt^2+d\rho^2,
\end{equation}
the \emph{proper momentum} is the momentum conjugate to $\rho$ along the worldline,
\begin{equation}
P_\rho \equiv \frac{\partial L}{\partial \dot\rho}
\quad\text{(with $L=-m\sqrt{-g_{\mu\nu}\dot x^\mu\dot x^\nu}$)}.
\end{equation}
The holographic dictionary entry proposed and verified in their AdS$_3$/CFT$_2$ examples is
\begin{equation}
\boxed{\;\dot{\mathcal{C}}(t)\sim P_\rho(t)\;}\label{caputadefi}
\end{equation}
(up to the normalization conventions).

In summary, the procedure is this: first, compute $\dot{\mathcal{C}}(t)$ from CFT correlators via the Krylov chain. Second,
calculate $P_\rho(t)$ from the infalling geodesic in proper-distance gauge and finally  match them.
\\
In what follows, we apply this logic for our AdS$_7$ system. It is important that in our case, the strongly coupled system is not described in terms of a quiver field theory. The best (and only) description is in terms of the holographic background in eq.(\ref{backgroundads7xm3}). The application of the recipe of \cite{Caputa:2024sux} to our case is not accompanied by a CFT calculation, though the ingredients are present. We allow the massive particle to move  along the AdS-radial coordinate, the $S^2(\theta,\varphi)$ directions and the quiver $\eta$-direction. In this way the massive particle falling along a geodesic represents the time evolution of an operator of the CFT$_6$. This operator has SU(2) R-charge (enacted by the motion on $S^2$), and spreads along the different nodes in the quiver, represented by the motion along $\eta$-coordinate. The treatment is reminiscent of that in \cite{Fatemiabhari:2025poq}, with the added subtlety of the R-symmetry quantum numbers.

Let us start studying a generic geodesic.

\subsection{A general geodesic}

We consider a time-like geodesic on the family of backgrounds described by eqs.(\ref{backgroundads7xm3})-(\ref{alpha}). The geodesics is parametrised by the $t$-coordinate. We propose the following (consistent) embedding,
\begin{equation}
r(t),~~\theta(t)=\frac{\pi}{2},~~\varphi(t),~~\eta(t). \label{embedding}
\end{equation}
The induced metric along the geodesic is
\begin{equation}
 ds_{ind,E}^2= \Bigg[f_1(\eta)\left( - e^{-2r} + \dot{r}^2\right) + f_2(\eta)\dot{\eta}^2 + f_3(\eta)\dot{\varphi}^2\Bigg] dt^2,
\end{equation}
and the action associated with this particle is
\begin{equation}
    S = -m\int dt \, \sqrt{f_{1}(\eta) \left(e^{-2r}-\dot{r}^{2}\right) -f_{2}(\eta)\dot{\eta}^{2}-f_{3}(\eta)\dot{\varphi}^{2}} .
\end{equation}
We also define  the {Lagrangian} density along the worldline
\begin{equation}
    L = \sqrt{f_{1}(\eta) \left(e^{-2r}-\dot{r}^{2}\right) -f_{2}(\eta)\dot{\eta}^{2}-f_{3}(\eta)\dot{\varphi}^{2}},
\end{equation}
so that the particle action reads $S=-m \int dt L$. Varying respect to $r,\eta$ and $\varphi$
leads to the following equations of motion,
\be
\frac{d}{dt}\left(\frac{f_{1}(\eta)}{L}\dot{r}\right) = \frac{f_{1}(\eta)}{L}e^{-2r},\label{eqr}
\ee
\begin{equation}
    \frac{d}{dt}\left(\frac{f_{2}(\eta)}{L}\dot{\eta}\right) = -\frac{1}{2L}\Bigg[f_{1}'(\eta)(e^{-2r}-\dot{r}^{2})-f_{2}'(\eta)\dot{\eta}^{2}-f_{3}'(\eta)\dot{\varphi}^{2}\Bigg],\label{eqeta}
\end{equation}
\begin{equation}
    \frac{d}{dt}\left[ \frac{f_3(\eta) ~\dot{\varphi}}{ L}\right]=0.\label{eqvarphi}
\end{equation}
We have two conserved quantities in this system. One is the Hamiltonian, given by
\begin{equation}
     H = \frac{m f_{1}(\eta)e^{-2r}}{\sqrt{f_{1}(\eta) \left(e^{-2r}-\dot{r}^{2}\right) -f_{2}(\eta)\dot{\eta}^{2}-f_{3}(\eta)\dot{\varphi}^{2}}}.\label{hamiltonian}
\end{equation}
The second conserved quantity is the angular momentum $J$, from eq.(\ref{eqvarphi})
\begin{equation}
    J = \frac{f_{3}(\eta)}{L}\dot{\varphi}.\label{Jcons}
\end{equation}
Using the Hamiltonian in eq.(\ref{hamiltonian}) and the angular momentum in eq.(\ref{Jcons}) we solve for $\dot\eta$ and $\dot\varphi$,

\begin{eqnarray}
 & &\dot{\varphi}= \frac{m J}{H}\frac{f_1(\eta)}{f_3(\eta)}e^{-2r(t)}, \label{phidoretadot}\\
 & &\dot{\eta}=\pm \frac{e^{-2r(t)}}{H}\sqrt{\frac{f_1(\eta)}{f_2(\eta) f_3(\eta)}} \times \sqrt{e^{2r(t)} H^2 f_3(\eta) - m^2 f_1(\eta)( J^2 + f_3(\eta)) - H^2 e^{4r(t)} f_3(\eta) \dot{r}^2}.\nonumber 
\end{eqnarray}

Taking derivatives of these and replacing them in eqs.(\ref{eqr})-(\ref{eqvarphi}), one finds that all equations of motion are satisfied if,
\begin{equation}
 e^{2r(t)}\ddot{r} + 2 e^{2r(t)}\dot{r}^2 -1=0.  \label{rdet}
\end{equation}
Equation~(\ref{rdet}) is a nonlinear ODE for the radial coordinate $r(t)$
that decouples completely from $\eta$ and $\phi$.
One can verify by direct substitution that, subject to the initial
conditions $r(0)=r_{\mathrm{UV}}$ and $\dot r(0)=0$ (a particle
released from rest at the UV boundary), the unique solution is
\begin{equation}
r(t) = \tfrac{1}{2}\log\!\left(e^{2r_{\mathrm{UV}}}+t^2\right).\label{rsol}
\end{equation}
This solution describes a particle falling from $r=r_{\mathrm{UV}}$
towards larger values of $r$ (deeper in the bulk) and is precisely
the same as the geodesic found in the Poincar\'e patch of AdS$_3$
in reference~[12].  The agreement is not accidental: the
$(t,r)$-sector of our AdS$_7$ background in Poincar\'e coordinates
--see eq.~(\ref{backgroundads7xm3})-- takes the form
\begin{equation*}
ds^2\big|_{(t,r)} = f_1(\eta)\!\left(-e^{-2r}dt^2+dr^2\right).
\end{equation*}
After absorbing $f_1(\eta_0)$ into a rescaling of the mass
parameter, reduces to the two-dimensional Poincar\'e AdS$_2$ metric
analysed in \cite{Caputa:2024sux}.  It is therefore the Poincar\'e patch, not the
global patch, that governs the radial dynamics throughout this paper.
The distinction matters at the level of boundary conditions and of
the late-time $r\to\infty$ asymptotics, which are those of a
Poincar\'e horizon.
%
%

Replacing this in the expressions in eq.(\ref{phidoretadot}), we find

\begin{eqnarray}
& &    \dot{\varphi} = \frac{J}{H}\frac{f_{1}(\eta)}{f_{3}(\eta)}\Big(\frac{1}{t^{2}+e^{2r_{UV}}}\Big),\label{etadotphidot}\\  
& & 
    \dot{\eta} = \pm \frac{1}{H}\Big(\frac{1}{e^{2r_{UV}}+t^{2}}\Big) \sqrt{\frac{f_{1}(\eta)}{f_{2}(\eta)f_{3}(\eta)}\left[H^{2}e^{2r_{UV}}f_{3}(\eta)-m^{2}f_{1}(\eta)(J^{2}+f_{3}(\eta) )\right]}.\nonumber
\end{eqnarray}
The problem to be solved is the following: given a 6d ${\cal N}=(1,0)$ SCFT characterised by a function $\alpha(\eta)$--and its dual background described by the functions $f_i(\eta)$ in eq.(\ref{functionsf})-- we need to solve for $\eta(t), \varphi(t)$ in eqs.(\ref{etadotphidot}). The solutions to these equations describe the geodesic in eq.(\ref{embedding}), for the holographic dual background to a particular ${\cal N}=(1,0)$ superconformal field theory of the quiver type.
In terms of the SCFT, an operator with a given conformal dimension and R-charge spreads into other operators with the same R-charge (motion in $\varphi$ with conserved $J$) and into operators that connect with different nodes of the quiver (this is allowed by the presence of bifundamental fields). The motion along the radial coordinate, described by eqs.(\ref{rdet}),(\ref{rsol}) is the same as in \cite{Caputa:2024sux} and reminds us that the evolution/spread is constrained by conformal symmetry.
\\
In forthcoming sections, we discuss the resolutions of the equations
(\ref{etadotphidot}), for quivers 1 and 2, discussed in Section \ref{sectwoexamples}. Before that, it is useful to discuss the {\it proper momentum}, which generalises that of \cite{Caputa:2024sux}. We propose that the rate of change of the complexity is proportional to this {\it generalised proper momentum} in analogy with eq.(\ref{caputadefi}).
\subsection{Generalised proper momentum}\label{gen-prop-mom}
The key idea behind the generalised proper momentum is to promote the
single `radial' coordinate of reference \cite{Caputa:2024sux} to a composite coordinate
$\rho$ that captures all dynamically active directions of the bulk
geometry.  In reference \cite{Caputa:2024sux} the complexity growth rate was identified
with the proper radial momentum $P_r$ of a particle falling in an
asymptotically AdS$_3$ background.  In our setting the particle moves
simultaneously along the AdS$_7$ radial direction $r$, the quiver
direction $\eta$, and the $S^2$ angle $\phi$.  It is therefore natural
to introduce a single arc-length parameter $\rho$ along the geodesic
that accounts for all three contributions.  Concretely, we require
$d\rho^2$ to be the line element induced on the three-dimensional
submanifold $(r,\eta,\phi)$ by the background metric, weighted by the
same conformal factors $f_i(\eta)$ that appear in the equations of
motion:
\[
d\rho^2 \;=\; f_1(\eta)\,dr^2 + f_2(\eta)\,d\eta^2 + f_3(\eta)\,d\phi^2.
\]
The momentum conjugate to this composite arc-length then receives
contributions from the individual canonical momenta $P_r$, $P_\eta$,
$P_\phi$ weighted by the Jacobians $\partial\dot r/\partial\dot\rho$,
$\partial\dot\eta/\partial\dot\rho$, $\partial\dot\phi/\partial\dot\rho$.
These Jacobians are precisely the ratios of the proper-distance
elements computed below, so the final result does read as we derive below--see eq.(\ref{generalisedproper}).
This is the natural multi-dimensional generalisation of the proper radial
momentum of \cite{Caputa:2024sux}.  This construction was introduced in \cite{Fatemiabhari:2025poq, Fatemiabhari:2026goj} and we
refer the reader there for additional details; here we have presented
the logic that motivates it in the context of the present system.

In more detail, we start defining the usual canonical momenta, as derivatives of the Lagrangian density with respect to the generalised velocity.
We have
\begin{eqnarray}
& & P_r= \frac{m f_1(\eta) \dot{r}}{\sqrt{f_1(\eta)\left( e^{-2r(t)} -\dot{r}^2\right) - f_2(\eta)\dot{\eta}^2 - f_3(\eta)\dot{\varphi}^2 }}, \nonumber\\
&& P_\eta=  \frac{m f_2(\eta) \dot{\eta}}{\sqrt{f_1(\eta)\left( e^{-2r(t)} -\dot{r}^2\right) - f_2(\eta)\dot{\eta}^2 - f_3(\eta)\dot{\varphi}^2 }}, \nonumber\\
& & P_\varphi=\frac{m f_3(\eta) \dot{\varphi}}{\sqrt{f_1(\eta)\left( e^{-2r(t)} -\dot{r}^2\right) - f_2(\eta)\dot{\eta}^2 - f_3(\eta)\dot{\varphi}^2 }}.
\end{eqnarray}
We can use eq.(\ref{rdet}) for $r(t)$ together with the Hamiltonian in eq.(\ref{hamiltonian})  to find
\begin{equation}
    P_{r} = \frac{m ~~ t ~~f_1(\eta)} {\left( e^{2r_{UV} }+t^2\right) {\sqrt{\frac{e^{2r_{UV}}}{(e^{2r_{UV}}+t^{2})^{2}}f_{1}(\eta)-f_{2}(\eta)\dot{\eta}^{2}-f_{3}(\eta)\dot{\varphi}^{2}}}} =Ht .\label{Prfinal}
\end{equation}
We also find,
\begin{equation}
    P_{\eta} = \frac{mf_{2}(\eta)\dot{\eta}}{\sqrt{\frac{e^{2r_{UV}}}{(e^{2r_{UV}}+t^{2})^{2}}f_{1}(\eta)-f_{2}(\eta)\dot{\eta}^{2}-f_{3}(\eta)\dot{\varphi}^{2}}}, 
    \end{equation}
    \begin{equation}
    P_{\varphi} = \frac{mf_{3}(\eta)\dot{\varphi}}{\sqrt{\frac{e^{2r_{UV}}}{(e^{2r_{UV}}+t^{2})^{2}}f_{1}(\eta)-f_{2}(\eta)\dot{\eta}^{2}-f_{3}(\eta)\dot{\varphi}^{2}}} \, .
\end{equation}
The idea now is to define a `proper coordinate' $\rho$, such that
\begin{equation}
     d\rho^{2} = f_{1}(\eta)dr^{2}+f_{2}(\eta)d\eta^{2}+f_{3}(\eta)d\varphi^{2} \, .
\end{equation}
Following the treatment in \cite{Fatemiabhari:2025poq, Fatemiabhari:2026goj}, we encounter the set of equalities,
\begin{eqnarray}
& &     \mathcal{C}^{-1} \equiv \frac{d\rho}{d r} = \pm \sqrt{f_{1}(\eta)+f_{2}(\eta)\left(\frac{\dot{\eta}}{\dot{r}}\right)^{2}+f_{3}(\eta)\left( \frac{\dot{\varphi}}{\dot{r}}\right)^{2}}, \nonumber\\
& & 
    \mathcal{A}^{-1} \equiv \frac{d\rho}{d \eta} = \pm\sqrt{f_{1}(\eta)\left(\frac{\dot{r}}{\dot{\eta}}\right)^{2}+f_{2}(\eta)+f_{3}(\eta)\left( \frac{\dot{\varphi}}{\dot{\eta}}\right)^{2}}, \nonumber\\
& &    \mathcal{B}^{-1} \equiv \frac{d\rho}{d \varphi} = \pm\sqrt{f_{1}(\eta)\left(\frac{\dot{r}}{\dot{\varphi}}\right)^{2}+f_{2}(\eta)\left(\frac{\dot{\eta}}{\dot{\varphi}}\right)^{2}+f_{3}(\eta)}.\label{usefuleqs} 
\end{eqnarray}
With these, the {\it generalised proper momentum} is 
\begin{equation}
    \begin{split}
        P_{\rho} &= P_{r} \frac{\partial \dot{r}}{\partial \dot{\rho}} + P_{\eta}\frac{\partial\dot{\eta}}{\partial\dot{\rho}}+P_{\varphi}\frac{\partial\dot{\varphi}}{\partial\dot{\rho}} 
        = \mathcal{C} P_{r} + \mathcal{A}P_{\eta}+\mathcal{B}P_{\varphi}=\sqrt{\frac{H^2 \left(e^{2 r_{UV}}+t^2\right)}{f_{1}(\eta(t))}-1} \, .
    \end{split}\label{generalisedproper}
\end{equation}
In other words, the proper momentum receives a contribution from the usual canonical momentum along the radial direction, but also contributions from motion `across the quiver', represented by ${\cal A} P_\eta$ and the motion in the R-symmetry direction given by ${\cal B}P_\varphi$. The evolution preserves the Hamiltonian $H$ in eq.(\ref{hamiltonian}) and the R-charge $J$ in eq.(\ref{Jcons}). Another way of expressing this is that the operator that spreads across the Hilbert space (as indicated at the beginning of this section), can have conserved quantum numbers (like an R-charge) and also explore different nodes in the quiver (which are connected by bifundamental fields).

It is important to mention that the signs in eq.\eqref{usefuleqs} should be chosen appropriately according to the direction of motion along the geodesic. For instance, if a particle is falling from $r_{UV}\sim-\infty$  towards larger values of the $r$ coordinate, while in $\rho$ coordinate the motion is mapped to a trajectory from $\rho_{UV}\sim+\infty$ to smaller values of $\rho$, a minus sign should be chosen for the first relation in eq.\eqref{usefuleqs}.  For $\eta$ and $\varphi$ directions, the same applies. In all cases, the last relation in eq. \eqref{generalisedproper} remains valid and does not need further adjustments. 

Finally, we {\it propose} that the relation with the rate of change of the complexity is that in eq.(\ref{caputadefi}), but using the generalised proper momentum. Namely
\begin{equation}
    \dot{C}(t) \sim P_{\rho} \, .
\end{equation}
In what follows, we solve the equations of motion (\ref{etadotphidot}) for the quivers 1 and 2 in Section \ref{sectwoexamples}, as case studies. We need to resort to numerical study, which otherwise is simple.
We study first the two quivers in the case of zero angular momentum $J=0$ (this is the operator that spreads has zero R-charge). This allows us to understand how the operator spreads across the quiver direction (an operator made out of bifundamental fields, it will spread across different nodes). After this, we discuss the effect of adding angular momentum (R-charge, $J\neq 0$) to this dynamics.

\subsection{Geodesics with $J=0$}\label{sec:J0}
In this section, we study the motion of the particle for the case of zero angular momentum $J = 0$. We make generic choices of initial position $\eta_{0} \equiv \eta(t = 0)$ along with vanishing initial velocity in the quiver direction $\dot{\eta}(t= 0) = 0$. To have non-trivial $\eta$-dynamics, we must choose $\eta_{0}$ such that $f_{1}'(\eta = \eta_{0}) \neq 0$, as can be seen from  equation (\ref{eqeta}). Recall that $J\sim \dot{\varphi}=0$ is assumed in this section. \\
For this case, the Hamiltonian in equation (\ref{hamiltonian}) simplifies to, 
\begin{equation}
    H=\frac{m\sqrt{f_{1}(\eta_{0})}}{e^{r_{UV}}} \, .
\end{equation}
Substituting this expression in the second equation of (\ref{etadotphidot}), one gets
\begin{equation}
     \dot{\eta} = \pm\frac{e^{r_{UV}}}{t^{2}+e^{2r_{UV}}}\sqrt{\frac{f_{1}(\eta)}{f_{2}(\eta)}\left(1-\frac{f_{1}(\eta)}{f_{1}(\eta_{0})}\right)} \, ,
     \label{eqeta0}
\end{equation}
which is consistent with the choice of initial conditions. Eq.(\ref{eqeta0}) implies 
\begin{equation}
    f_{1}(\eta_{0}) \geq f_{1}(\eta(t)) \, .
\end{equation}

\begin{figure}[]
    \centering
    \includegraphics[width=0.4\linewidth]{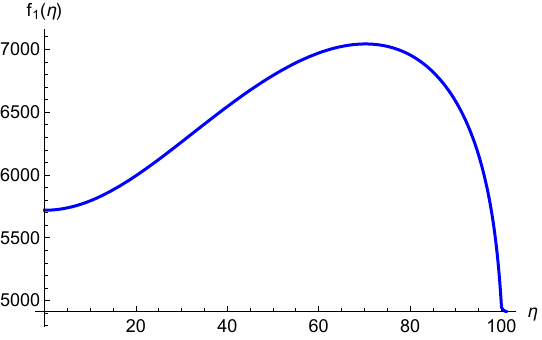}
    \quad \quad \quad 
    \includegraphics[width=0.4\linewidth]{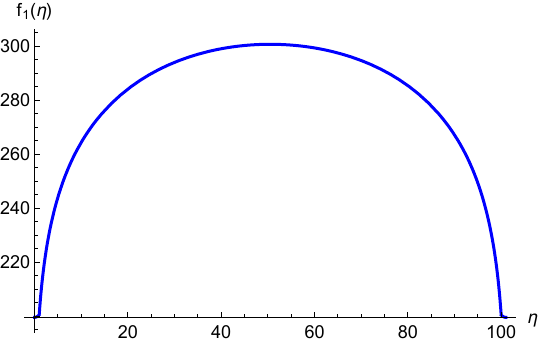}
    \caption{Plot of $f_{1}(\eta)$ for both quiver 1 (left) and quiver 2 (right). We set $P = 100$.
    }
    \label{f1}
\end{figure}

Additionally, eq.(\ref{eqeta0}) allows solutions with negative and positive velocity. The sign of the velocity is fixed by the choice of initial position in the $\eta$-coordinate. By means of eq.(\ref{eqeta}), it is possible to understand how the sign is fixed by treating $f_{1}(\eta)$ as a potential term. An interesting feature is that, for both of the quivers, the function $f_{1}(\eta)$  has a global maximum located at $\eta_{\text{max}}$ (see Figure \ref{f1}). As a consequence, the point $\eta = \eta_{\text{max}}$ acts as an equilibrium position for the particle, meaning that if the particle starts at $\eta_{0} =\eta_{\text{max}} $, it remains there for all times.

From the dual field theory perspective, this equilibrium point
corresponds to a specific gauge node of the quiver at which an
operator can reside without spreading to neighbouring nodes.
To the best of our knowledge, the existence of such a preferred
node is a \emph{novel prediction} of the holographic analysis; it does not follow from any known field theory argument about
operator dynamics in these 6d SCFTs, whose Lagrangian description is
not available.  It would be very interesting to understand whether
this feature can be reproduced or anticipated from the structure of
the quiver (for example, from the rank function $R(\eta)$ or the balance
condition), if a direct field-theory calculation of Krylov
complexity becomes available.  For the time being, we take it as a
concrete, falsifiable prediction of the holographic complexity
proposal.
\\
For the two  examples discussed in this paper, it is possible to compute $\eta_{\text{max}}$ in terms of the quiver parameters. \\
From a field theory perspective, this indicates that if the operator starts localised at a specific gauge node of the quiver (associated with $\eta_{\text{max}}$), it will not spread to the rest of the nodes. 

Since the quiver is of finite  size (the $\eta$ coordinate has a finite range), the particle may reach the ends of the $\eta$-space. In these cases, the particle \textit{bounces} back elastically,  changing the sign of the velocity $\dot{\eta}(t_{*}) \to -\dot{\eta}(t_{*}) $. Here, $t_{*}$ is the time for which the particle reaches the end of the $\eta$-coordinate. Note that this sign change is allowed by eq. (\ref{eqeta0}).

\subsubsection{Quiver 1} \label{sec:J0Q1}
For this first example, we consider the quiver closed by a flavour node, described by the function $\alpha_{1}(\eta)$. See Section \ref{subsection:quiver1} for details.

\begin{figure}[]
    \centering
    \includegraphics[width=0.5\linewidth]{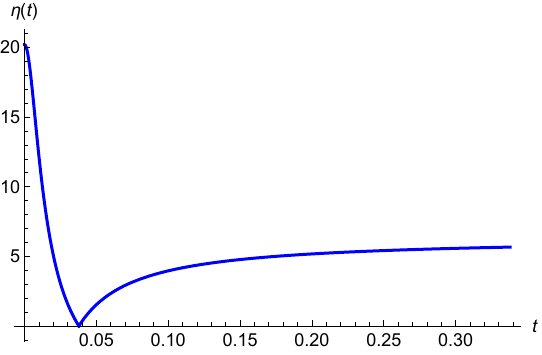}
    \caption{Trajectory of the particle in $\eta$ for quiver 1. We set $J = 0$, $\eta_{0} = 20$, $P = 100$ and $e^{r_{UV}} = 0.01$.}
    \label{quiver120eta}
\end{figure}

\begin{figure}
    \centering
    \includegraphics[width=0.4\linewidth]{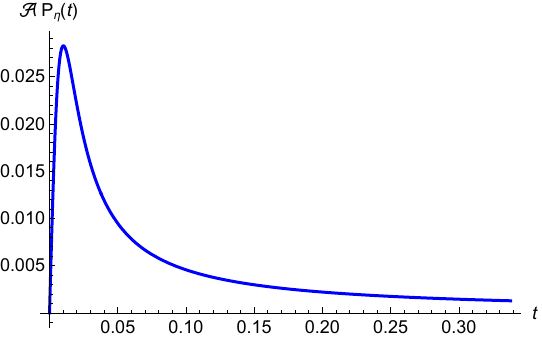}
    \quad \quad
    \includegraphics[width=0.4\linewidth]{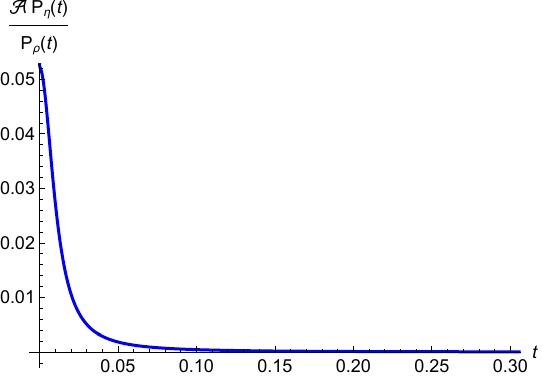}
    \caption{Momentum of the particle along the $\eta$ direction (left) and contribution of the $\eta$ momentum to the proper radial momentum. We set $J = 0$, $\eta_{0} = 20$, $P = 100$ and $e^{r_{UV}} = 0.01$.}
    \label{quiver120p}
\end{figure}

\begin{figure}[]
    \centering
    \includegraphics[width=0.5\linewidth]{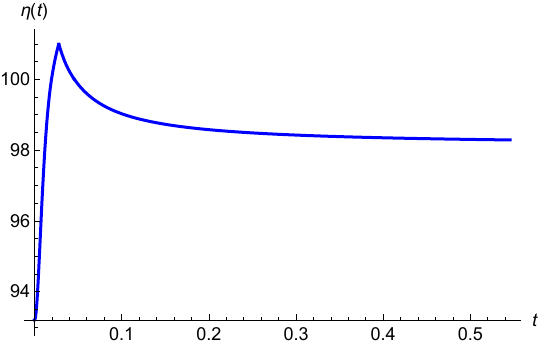}
    \caption{Trajectory of the particle in $\eta$ for quiver 1.  We set $J = 0$, $\eta_{0} = 93$, $P = 100$ and $e^{r_{UV}} = 0.01$.}
    \label{quiver193}
\end{figure}

The full numerical solution for $\eta(t)$ is shown in Figures \ref{quiver120eta} and \ref{quiver193}. For the first case, the particle moves towards decreasing values of $\eta$ until it reaches the end of the space (at $\eta=0$ or at $\eta=P+1$), from which it bounces back and quickly stabilises. The motion on $\eta$ is damped as the Figures \ref{quiver120eta} and \ref{quiver193} show. The contributions to the proper momentum from the $\eta$-momentum $P_\eta$, are relevant at early times (see Figure \ref{quiver120p}). On the other hand, for large $t$, equation (\ref{generalisedproper}) becomes
\begin{eqnarray}
    P_{\rho}(t)\bigg|_{t \to \infty} = H t = P_{r} \, ,
    \label{prholarget}
\end{eqnarray}
meaning that for large times, the motion along the radial direction of AdS dominates the dynamics. This is in agreement with the linear growth of complexity and the fact that the dual theory is conformal. 

\begin{figure}
    \centering
    \includegraphics[width=0.4\linewidth]{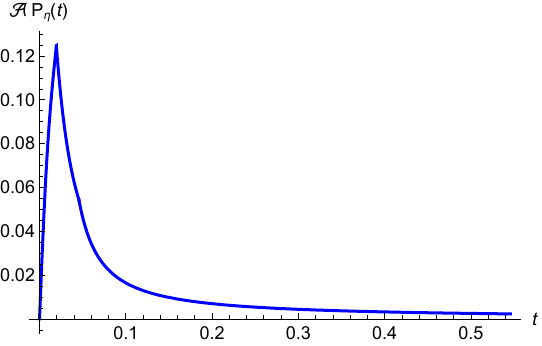}
    \quad \quad 
    \includegraphics[width=0.4\linewidth]{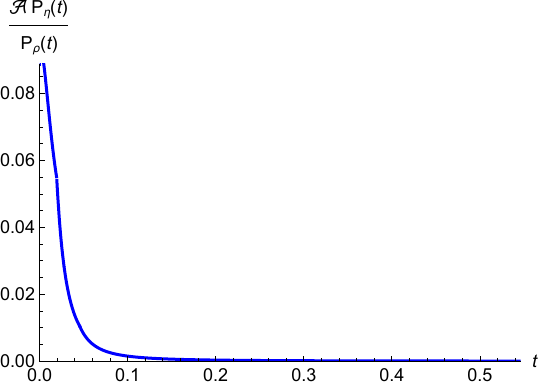}
    \caption{Momentum of the particle along the $\eta$ direction (left) and contribution of the $\eta$ momentum to the proper radial momentum. We set $J = 0$, $\eta_{0} = 93$, $P = 100$ and $e^{r_{UV}} = 0.01$.}
    \label{quiver193p}
\end{figure}

The second case is depicted in Figures \ref{quiver193} and \ref{quiver193p}. Here, we consider a particle starting in the first branch of the quiver ( for $0 \leq\eta\leq P$) and near the final gauge node of the quiver. In this case, the particle moves towards increasing $\eta$, reaching the final gauge node located at $\eta = P$. Then, it continues to the second branch of the quiver ($P \leq \eta \leq P+1$) towards the flavour node at $\eta = P+1$. Finally, the particle returns by changing the sign of the $\eta$-velocity in eq.(\ref{eqeta0}) and returning to the first branch of the quiver. Despite the change in the direction of motion in $\eta$, the motion is still damped.  Figure \ref{quiver193p} shows that the contribution of the $\eta$ motion to the proper  momentum, is still localised at early times, in agreement with equation (\ref{prholarget}).
\\
Before proceeding to the second quiver, it is instructive to
identify the timescale that separates the `early-time' quiver
dynamics from the `late-time' AdS-radial regime.
Inspecting eq.~(\ref{eqeta0}), the $\eta$-velocity is controlled by the
factor $e^{r_{\text{UV}}}(e^{2r_{\mathrm{UV}}}+t^2)^{-1}$, which becomes
$\mathcal{O}(1)$ when $t\sim e^{\frac{r_{\mathrm{UV}}}{2}}$.  Hence
the characteristic crossover time is
\[
t_{cross} \;\sim\; e^{\frac{r_{\mathrm{UV}}}{2}}.
\]
This crossover time is set by the UV boundary position in Poincar\'e coordinates.
In the AdS/CFT dictionary, $e^{r_{\mathrm{UV}}}$ is 
proportional to the UV cutoff $\epsilon$ of the boundary theory (with
$\epsilon\to 0$ corresponding to $r_{\mathrm{UV}}\to-\infty$ in our
conventions.
The AdS radius $L_{\mathrm{AdS}}$, which is absorbed into the
functions $f_i(\eta)$, sets the overall scale for the momentum
but does not independently modify $t_{cross}$.  At early times
$t<t_{cross}$, the $\eta$- and $\phi$-momenta can be comparable
to $P_r$ and the quiver/R-charge dynamics is visible; for
$t\gg t_{cross}$ the factor $(e^{2r_{\mathrm{UV}}}+t^2)^{-1}$ strongly
suppresses both $\dot\eta$ and $\dot\phi$, and $P_r=Ht$ dominates.
The quiver parameters (the integer $P$ and the choice of initial
node $\eta_0$) do not change $t_{cross}$ but they do determine the
\emph{amplitude} of the transient quiver dynamics, because they
control $f_1(\eta)$ and hence the depth of the effective potential
through which the particle moves.  For initial positions close to
the equilibrium $\eta_{\max}$, the amplitude of
the $\eta$-oscillations is small and the quiver contribution to
$P_\rho$ is already subdominant even at $t<t_{cross}$.

\subsubsection{Quiver 2} \label{sec:J0Q2}
In this example, we study the quiver in Section \ref{subsubsection:quiver2}. This has two flavour nodes, at the beginning and at the end of the quiver, with constant rank gauge nodes. This is holographically described  by the function $\alpha_{2}(\eta)$.

\begin{figure}[]
    \centering
    \includegraphics[width=0.5\linewidth]{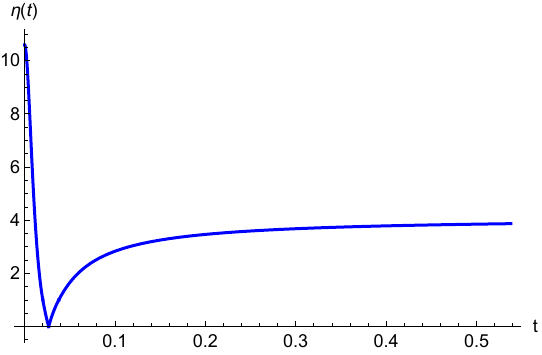}
    \caption{Trajectory of the particle in $\eta$ for quiver 2.  We set $J = 0$, $\eta_{0} = 10$, $P = 100$ and $e^{r_{UV}} = 0.01$.}
    \label{quiver21}
\end{figure}

\begin{figure}[]
    \centering
    \includegraphics[width=0.4\linewidth]{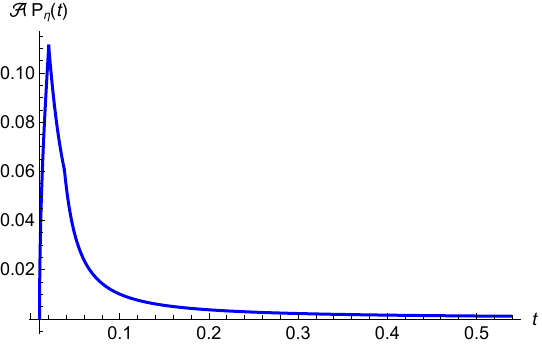}
    \quad \quad 
    \includegraphics[width=0.4\linewidth]{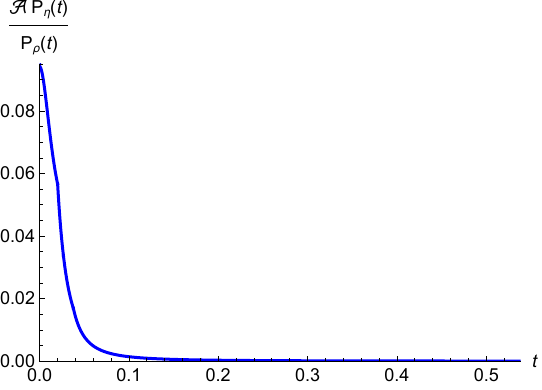}
    \caption{Momentum of the particle along the $\eta$ direction (left) and contribution of the $\eta$ momentum to the proper radial momentum. We set $J = 0$, $\eta_{0} = 10$, $P = 100$ and $e^{r_{UV}} = 0.01$.}
    \label{quiver21p}
\end{figure}

\begin{figure}[]
    \centering
    \includegraphics[width=0.5\linewidth]{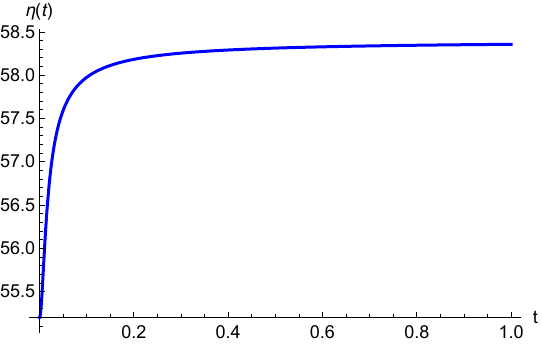}
    \caption{Trajectory of the particle in $\eta$ for quiver 2. We set $J = 0$, $\eta_{0} = 55$, $P = 100$ and $e^{r_{UV}} = 0.01$.}
    \label{quiver255}
\end{figure}

\begin{figure}[]
    \centering
    \includegraphics[width=0.4\linewidth]{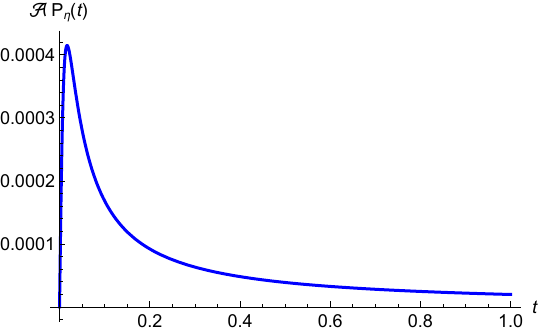}
    \quad \quad 
    \includegraphics[width=0.4\linewidth]{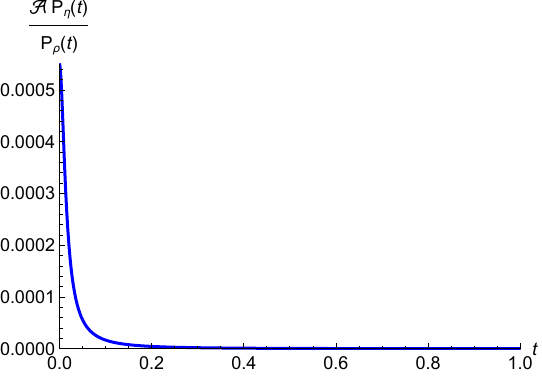}
    \caption{Momentum of the particle along the $\eta$ direction (left) and contribution of the $\eta$ momentum to the proper radial momentum for quiver 2. We set $J = 0$, $\eta_{0} = 55$, $P = 100$ and $e^{r_{UV}} = 0.01$.}
    \label{quiver255p}
\end{figure}

The first trajectory we study is shown in Figure \ref{quiver21}. The particle starts at the second branch of the quiver ($1\leq \eta \leq P$), quickly moves towards decreasing $\eta$, and changes to the first branch ($0\leq \eta \leq 1$) followed by a bounce when reaching $\eta = 0$. After this, it bounces back on the first branch in the opposite direction, reaching the second branch, where it stops. The $\eta$-momentum of the particle is shown in Figure \ref{quiver21p}. The behavior is analogous to the one discussed in the previous example: at large times, the behavior is a generic CFT--see eq.(\ref{prholarget}). \\
The second case study is shown in Figures \ref{quiver255} and \ref{quiver255p}, unlike the previous cases, the initial position of the particle is chosen such that the particle explores the middle section of the quiver without reaching the end-points. In this case, the particle goes toward increasing $\eta$, and explores only the second branch of the quiver, meaning $1 \leq \eta(t) \leq P$, quickly stopping at a neighboring gauge node. Overall, the behavior agrees with equation (\ref{prholarget}) at large times, as both of the quivers describe CFTs. \\

\textbf{Summary}: This section was devoted to the study of geodesics with zero angular momentum $J = 0$, and how the particle explores the quiver direction by choosing different initial positions. One important aspect to focus on is the existence of an equilibrium position $\eta_{\max}$.  From the dual field-theory perspective this indicates
that a CFT operator initially located at the corresponding gauge node
does not spread along the quiver.  We stress that this equilibrium
is a genuinely new prediction of the holographic approach. There is
currently no known field-theory mechanism or expectation that
pinpoints such a preferred node.  In addition, we have shown that
the representative choices of initial conditions explored here, 
initial positions near the centre of the quiver, near the boundary,
and near $\eta_{\max}$, exhibit qualitatively distinct early-time
behaviour (different amplitudes and directions of $\eta$-motion,
different contributions of $P_\eta$ to $P_\rho$) but share the same
universal CFT late-time asymptotics, eq.~(\ref{prholarget}).  We have checked that
choosing other initial positions within each branch of the quiver does
not reveal qualitatively new phenomena. The dynamics always settles
into one of the three representative regimes described above.
Moreover, we also highlight the CFT behaviour at large times, which
agrees with the quadratic growth of the complexity.

%
%
%
%

\subsection{Geodesics with $J\neq 0$} \label{sec:Jn0}
In this section, we study the motion of the particle for the case of non-zero angular momentum $J \neq 0$. We make generic choices of initial position $\eta_{0} \equiv \eta(t = 0)$ along with zero  initial velocity in the quiver direction $\dot{\eta}(t= 0)\equiv \dot{\eta}_0 = 0$,  but with $\dot{\varphi}(t= 0)\equiv \dot{\varphi}_0 \neq 0$. 

The equations of motion to be solved are those in (\ref{etadotphidot}), which we quote here to ease the reading, 
\begin{eqnarray}
& &    \dot{\varphi} = \frac{J}{H}\frac{f_{1}(\eta)}{f_{3}(\eta)}\Big(\frac{1}{t^{2}+e^{2r_{UV}}}\Big),\\  
& & 
    \dot{\eta} = \pm \frac{1}{H}\Big(\frac{1}{e^{2r_{UV}}+t^{2}}\Big) \sqrt{\frac{f_{1}(\eta)}{f_{2}(\eta)f_{3}(\eta)}\left[H^{2}e^{2r_{UV}}f_{3}(\eta)-m^{2}f_{1}(\eta)(J^{2}+f_{3}(\eta) )\right]}.\label{etadotphidot2}
\end{eqnarray}
For this case, the Hamiltonian in eq.~(\ref{hamiltonian}) and the angular momentum in eq.~\eqref{Jcons} simplify to 
\begin{equation}
     H = \frac{m f_{1}(\eta_0)e^{-2r_{UV}}}{\sqrt{f_{1}(\eta_0) \left(e^{-2r_{UV}}\right) -f_{3}(\eta_0)\dot{\varphi}_0^{2}}}.\label{hamiltonian2}
\end{equation}
\begin{equation}
    J = \frac{f_{3}(\eta_0)}{\sqrt{f_{1}(\eta_0) \left(e^{-2r_{UV}}\right) -f_{3}(\eta_0)\dot{\varphi}_0^{2}}}\dot{\varphi}_0.\label{Jcons2}
\end{equation}

After using constraints in eqs.~\eqref{hamiltonian2}-\eqref{Jcons2}, analyzing  eq.~\eqref{etadotphidot2} shows that the function under the square root has one real root corresponding to the $(r_{UV},\eta_0)$ initial position and may have other real roots which restrict the range of the motion in some cases. This is due to the conservation of angular momentum, which is analysed on a case-by-case basis as follows in the next subsections.

As mentioned, in calculating the components of the generalised proper momenta defined in Section \ref{gen-prop-mom} as, 
\begin{equation}
    \begin{split}
        P_{\rho} &= P_{r} \frac{\partial \dot{r}}{\partial \dot{\rho}} + P_{\eta}\frac{\partial\dot{\eta}}{\partial\dot{\rho}}+P_{\varphi}\frac{\partial\dot{\varphi}}{\partial\dot{\rho}} 
        = \mathcal{C} P_{r} + \mathcal{A}P_{\eta}+\mathcal{B}P_{\varphi}=\sqrt{\frac{H^2 \left(e^{2 r_{UV}}+t^2\right)}{f_{1}(\eta(t))}-1},
    \end{split}\label{generalisedproper1}
\end{equation}
we should be cautious regarding the signs appearing in the following equations
\begin{eqnarray}
& &     \mathcal{C}^{-1} \equiv \frac{d\rho}{d r} = \pm \sqrt{f_{1}(\eta)+f_{2}(\eta)\left(\frac{\dot{\eta}}{\dot{r}}\right)^{2}+f_{3}(\eta)\left( \frac{\dot{\varphi}}{\dot{r}}\right)^{2}}, \label{usefuleqs1} \\
& & 
    \mathcal{A}^{-1} \equiv \frac{d\rho}{d \eta} = \pm\sqrt{f_{1}(\eta)\left(\frac{\dot{r}}{\dot{\eta}}\right)^{2}+f_{2}(\eta)+f_{3}(\eta)\left( \frac{\dot{\varphi}}{\dot{\eta}}\right)^{2}}, \label{usefuleqs2} \\
& &    \mathcal{B}^{-1} \equiv \frac{d\rho}{d \varphi} = \pm\sqrt{f_{1}(\eta)\left(\frac{\dot{r}}{\dot{\varphi}}\right)^{2}+f_{2}(\eta)\left(\frac{\dot{\eta}}{\dot{\varphi}}\right)^{2}+f_{3}(\eta)}.\label{usefuleqs3} 
\end{eqnarray}
In our analysis, the value of the $\rho$ coordinate is always decreasing from $\rho_{UV}$ to smaller values. On the other hand,  $r$ increases from $r_{UV}$ to larger values and $\varphi$ increases from zero to larger values. Hence, a negative sign should be picked for the eq. \eqref{usefuleqs1} while eq. \eqref{usefuleqs3} carries only a plus sign.  For the $\eta$ direction, the sign should be chosen depending on the direction of motion. 

In particular, if a bounce happens, this causes the particle to change the direction of motion, and then the sign changes. In all cases, the last relation in eq. \eqref{generalisedproper1} remains valid and does not need further adjustments. 

Let us now discuss the numerical results for the quivers 1 and 2 in Section \ref{sectwoexamples}.
\subsubsection{Quiver 1} \label{sec:Jn0Q1}
As the first example, we consider the quiver closed by a flavour node, described by the function $\alpha_{1}(\eta)$ defined in Section \ref{subsection:quiver1}.

\begin{figure}[]
    \centering
    \includegraphics[width=0.32\linewidth]{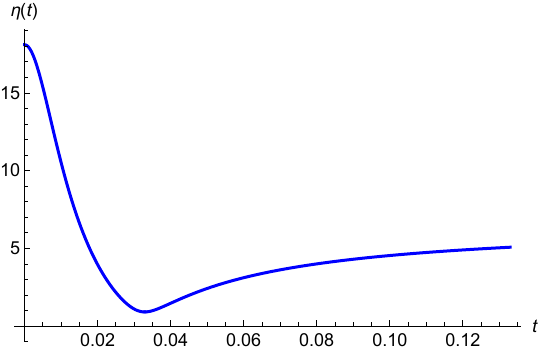}
    \includegraphics[width=0.32\linewidth]{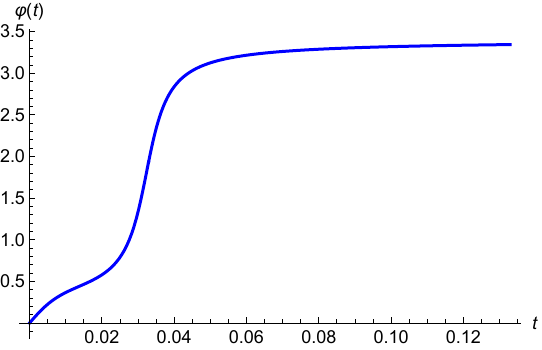}
    \caption{Trajectory of the particle in $\eta$ (left) and $\varphi$ (right) directions for Quiver 1 with $\eta_{0}= 18$,  $J = 0.0125$, $\dot{\varphi}_{0} = 50$,$P = 100$ and $e^{r_{UV}} = 0.01$.} \label{fig:Q1J}
\end{figure}

\begin{figure}[]
    \centering
    \includegraphics[width=0.32\linewidth]{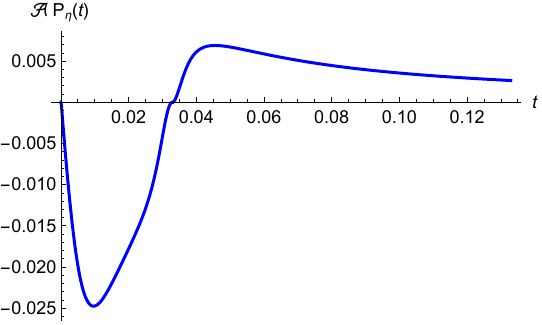}
    \includegraphics[width=0.32\linewidth]{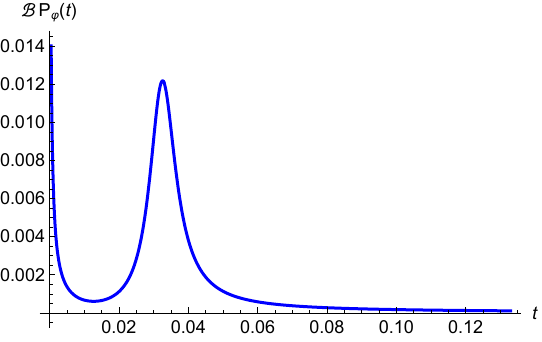}
    \includegraphics[width=0.32\linewidth]{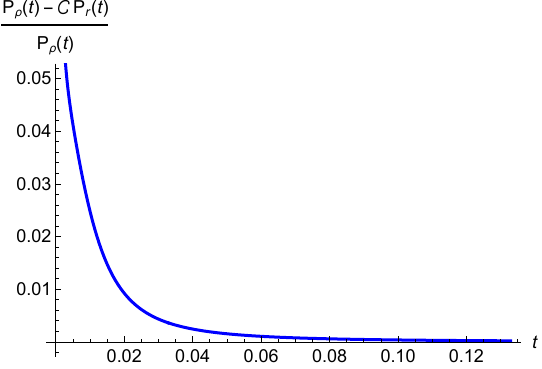}
    \caption{The components of the particle momentum in $\eta$ (left) and $\varphi$ (middle) directions and $(P_\rho-{\cal C}P_r)/P_\rho$ (right) for Quiver 1 with $\eta_{0}= 18$,  $J = 0.0125$, $\dot{\varphi}_{0} = 50$, $P = 100$ and $e^{r_{UV}} = 0.01$.} \label{fig:Q11J}
\end{figure}

The full numerical solution for $\eta(t)$ is shown in Figures \ref{fig:Q1J} and \ref{fig:Q11J}. In the first case, the particle moves towards decreasing values of $\eta$, but contrary to the $J=0$ case, it does not reach the end of the space (at $\eta=0$). It bounces back from a certain value of $\eta$ and quickly stabilises. This is due to the conservation of angular momentum, preventing the particle from reaching the end of space. It is clear from Figure \ref{fig:Q1J} that the momentum $P_\varphi$ (which is proportional to  $\dot \varphi$) increases as the particle approaches the turning point. At the same time, $\dot \eta$ is decreasing. The increase in the angular $\varphi$ momentum contributes in such a way that the constraints from the Hamiltonian and total angular momentum J are satisfied. After the bounce, the momentum in the $\eta$ direction increases again and takes over while angular rotation almost stops --- see first two panels in Figure \ref{fig:Q11J}. In other words, the particle can not reach $\eta=0$ otherwise the momentum in  $\varphi$ direction would diverge. 

The motions on the $\eta$ and $\varphi$ directions are damped as  Figure \ref{fig:Q1J} shows. The contributions to the proper momentum coming from $P_\eta$ and that coming from $P_\varphi$, are relevant at early times (see the last panel of Figure \ref{fig:Q11J}). On the other hand, for large $t$, equation (\ref{generalisedproper}) approximates
\begin{eqnarray}
    P_{\rho}(t)\bigg|_{t \to \infty} = H t = P_{r} \, ,
    \label{prholate}
\end{eqnarray}
This means that for late times, the motion along the radial direction of AdS dominates the dynamics. This is in agreement with the linear growth of complexity and the fact that the dual theory is conformal. We do not plot $P_{\rho}(t)$ for these examples as the behavior is linear and similar at the late times.

\begin{figure}[]
    \centering
    \includegraphics[width=0.32\linewidth]{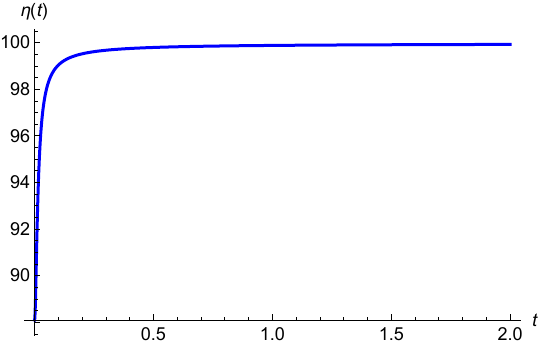}
    \includegraphics[width=0.32\linewidth]{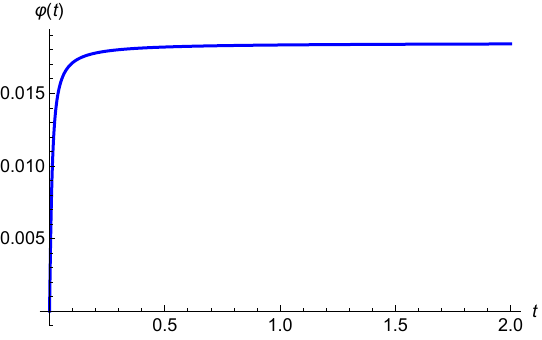}
    \includegraphics[width=0.32\linewidth]{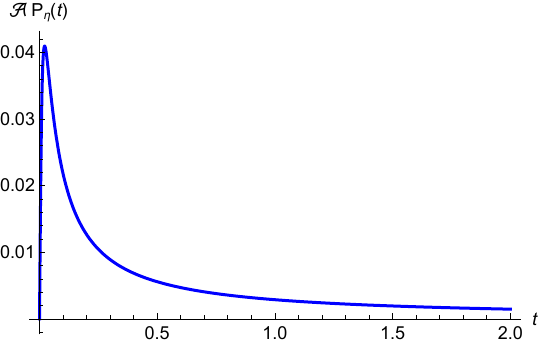}
    \includegraphics[width=0.32\linewidth]{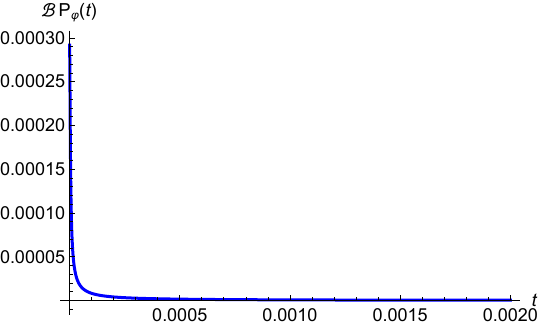}
    \includegraphics[width=0.32\linewidth]{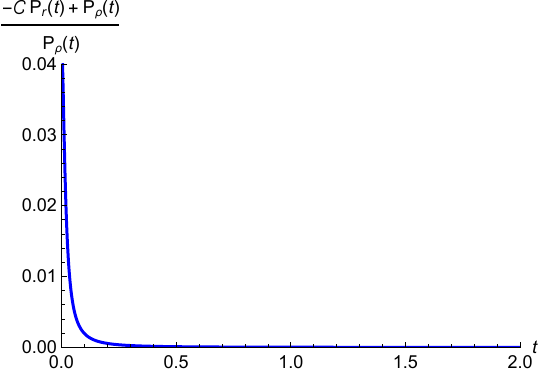}
    \caption{Trajectory of the particle in $\eta$ and $\varphi$ directions, the components of the particle momentum in $\eta$ and $\varphi$ directions and $(P_\rho-{\cal C}P_r)/P_\rho$ for Quiver 1 with $\eta_{0}= 88$,  $J = 6.67 \times 10^{-5}$, $\dot{\varphi}_{0} = 1.1$, $P = 100$ and $e^{r_{UV}} = 0.01$.} \label{fig:Q12J}
\end{figure}

A second situation is depicted in Figure \ref{fig:Q12J}. Here, the particle starts in the first branch of the quiver ( for $0 \leq\eta\leq P$) and near the final gauge node of the quiver. The particle moves towards increasing $\eta$, asymptotically reaching the final gauge node located at $\eta = P$. The motion is damped, and it does not reach the second branch of the quiver ($P \leq \eta \leq P+1$).  The contribution of the motion in $\eta$ and $\varphi$ directions to the proper momentum is still mostly localised at early times, in agreement with equation (\ref{prholate}). Let us now study the complexity for the second quiver.

\subsubsection{Quiver 2} \label{sec:Jn0Q2}
We study the quiver in Section \ref{subsubsection:quiver2}. This has two flavour nodes, at the beginning and at the end of the quiver, with constant rank gauge nodes. This information is encoded in the function $\alpha_{2}(\eta)$.

\begin{figure}[]
    \centering
    \includegraphics[width=0.32\linewidth]{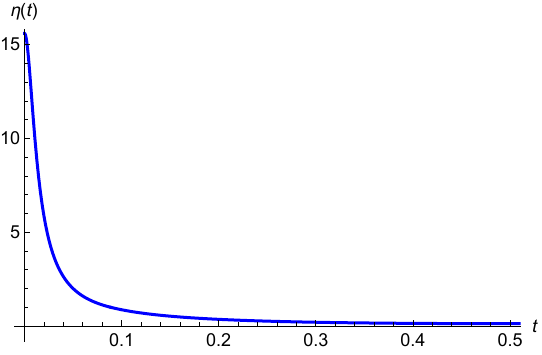}
    \includegraphics[width=0.32\linewidth]{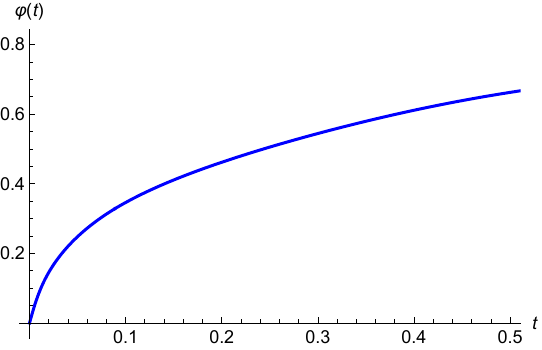}
    \caption{Trajectory of the particle in $\eta$ (left) and $\phi$ (right) directions for Quiver 2 with $\eta_{0}= 15$,  $J \neq 0$, $\dot{\varphi}_{0} = 10$,  $P = 100$ and $e^{r_{UV}} = 0.01$.} \label{fig:Q2J}
\end{figure}

\begin{figure}[]
    \centering
    \includegraphics[width=0.32\linewidth]{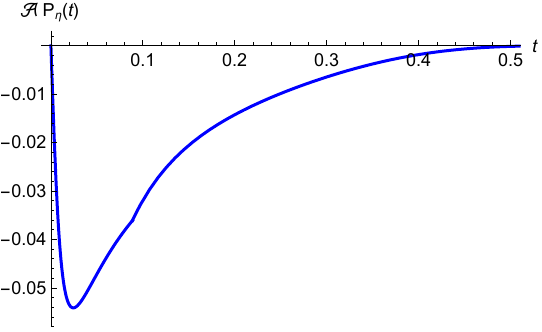}
    \includegraphics[width=0.32\linewidth]{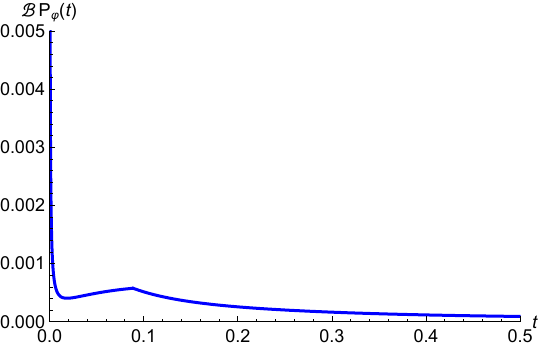}
    \includegraphics[width=0.32\linewidth]{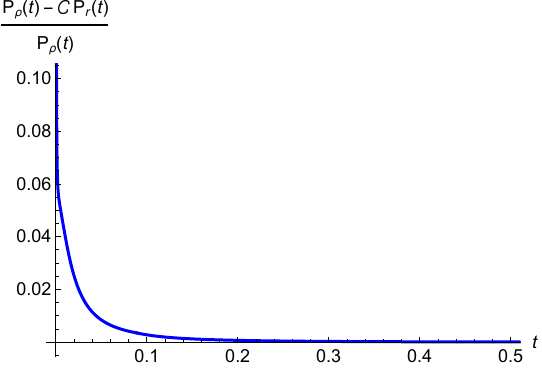}
    \caption{The components of the particle momentum in $\eta$ (left) and $\varphi$ (middle) directions and $(P_\rho-{\cal C}P_r)/P_\rho$ (right) for Quiver 2 with $\eta_{0}= 15$,  $J = 0.0001$, $\dot{\varphi}_{0} = 10$, $P = 100$ and $e^{r_{UV}} = 0.01$.} \label{fig:Q22J}
\end{figure}

The first trajectory we study is shown in Figure \ref{fig:Q2J}. The particle starts at the second branch of the quiver ($1\leq \eta \leq P$), quickly moves towards decreasing $\eta$, and changes to the first branch ($0\leq \eta \leq 1$), but it does not reach $\eta = 0$ and continues in the first branch where it stops. This should be compared with the similar study in Section \ref{sec:J0Q2} with zero $J$ as the particle bounces there. The $\eta$ and $\varphi$-momenta of the particle are shown in Figure \ref{fig:Q22J}. The behavior is analogous to the one discussed in the previous example, as at large times it resembles the one found in a generic CFT, see eq.(\ref{prholarget}). 

{\bf In summary}: in this section, we studied the geodesics of a particle falling in our background with non-zero angular momentum and different initial positions. 
The difference with the $J=0$ trajectory is caused by the angular-momentum term in the square root of eq.~(\ref{etadotphidot2}). This term acts as an effective centrifugal barrier in the internal space. As the particle moves towards the endpoint, part of the available energy is stored in the $\phi$-motion. The allowed region in $\eta$ is therefore smaller than in the zero-charge case, and the particle turns around before reaching the endpoint. This is the same mechanism observed for the first quiver, but in the second quiver it is particularly transparent because the middle region of the rank function is approximately flat. The effect is nevertheless transient. At late times the universal factor \((t^2+e^{2r_{\rm UV}})^{-1}\) suppresses the internal velocities, and the radial contribution dominates the generalized proper momentum.


\section{Discussion, Further Comments and Conclusions}\label{conclusionsection}
In this work, we studied aspects of Krylov (spread) complexity for a class of six-dimensional $\mathcal{N}=(1,0)$ superconformal field theories and their holographic duals in massive type IIA supergravity. These theories are engineered by Hanany-Witten brane constructions involving NS5, D6 and D8 branes and admit holographic descriptions in terms of $AdS_7$ backgrounds characterised by a function $\alpha(\eta)$ that encodes the data of the dual linear quiver. We reviewed this construction and presented two representative examples of quivers that we use as case studies. Building on recent proposals relating the growth rate of Krylov complexity to the proper momentum of an infalling particle in the bulk geometry, we extended this prescription to the $AdS_7$ backgrounds relevant for these six-dimensional theories.

Within this framework, the spreading of operators in the strongly coupled SCFT is modelled holographically by the motion of a massive particle following timelike geodesics in the bulk geometry. Motion along the internal $S^2$ directions encodes the presence of $SU(2)_R$ charge. This is in nice correspondence with symmetry resolved Krylov complexity \cite{Caputa:2025mii, Caputa:2025ozd}. Motion along the coordinate $\eta$ captures the possibility that the operator spreads across different nodes of the quiver. Solving the resulting equations of motion numerically for two illustrative quiver theories, both for vanishing and non-vanishing angular momentum, we found that the dynamics along the quiver direction is typically damped and largely restricted to early times, while at late times the motion is dominated by the $AdS$ radial direction. As a consequence, the generalised proper momentum, and hence the rate of holographic complexity growth, approaches a linear behaviour at late times, consistent with expectations for conformal theories. These results provide a geometric picture for how operator spreading, global charges and quiver structure are encoded in the holographic description of complexity in higher--dimensional SCFTs.
\\
It is worth commenting on the characteristic time scale
$t_{cross}\sim e^{\frac{r_{\mathrm{UV}}}{2}}$ that separates early- and late-time
behaviour in our geodesics.  This scale is set by the UV boundary
position and is thus a UV datum of the field theory rather than
a property of the bulk geometry per se.  In holographic terms
$e^{r_{\mathrm{UV}}}$ plays the role of the UV regulator, so as the
cutoff is removed ($e^{r_{\mathrm{UV}}}\to 0$) the transient quiver
window shrinks to $t_{cross}\to 0$, consistent with the expectation that,
in the strict CFT limit, operator spreading into the quiver nodes is
suppressed relative to the growth driven by the conformal radial
dynamics.  This also implies that the separation between quiver
and radial dynamics is \emph{not} a universal property of these
theories but depends on the choice of boundary conditions (initial
position $\eta_0$ and UV cutoff $e^{r_{\mathrm{UV}}}$).
\\
\\

Interestingly, one can make a comparison between the results with the non-zero angular momentum case studied here and similar results obtained in Ref. \cite{Fatemiabhari:2026goj}. In both cases, the proper momentum and complexity are altered considerably at early times, while in late time, the effect of the addition of angular momenta dual to the R-charge is less visible. It is noteworthy that in the case of this paper, late time behaviour of the complexity is dominated by that of CFTs. In contrast, for  confining backgrounds \cite{Fatemiabhari:2026goj, Fatemiabhari:2025usn}, it resembles the complexity growth in gapped systems \cite{Jiang:2025wpj}.

There are several interesting directions in which the present work could be extended. A natural development would be to perform the field theory calculation of Krylov complexity directly in the six-dimensional SCFTs discussed here, allowing for a more precise comparison with the holographic predictions. It would also be interesting to explore more general quiver configurations and study how the structure of the rank function $R(\eta)$ influences the spreading of operators across the quiver. Another promising direction is to analyse related observables in these backgrounds, such as operator growth, chaos diagnostics \cite{Nunez:2018qcj, Filippas:2019puw}, entanglement or other measures of quantum information, in order to better understand the interplay between complexity, global symmetries and the structure of strongly coupled higher-dimensional field theories. 
A natural companion question concerns the class of operators
whose complexity growth is captured by the present geodesic
analysis.  As noted in the Introduction, the point-particle
approximation is valid for single-trace primary operators of large
conformal dimension $\Delta\gg 1$ and fixed $R$-charge $J$.  For
operators that correspond to extended bulk objects, such as the
instanton strings, the appropriate bulk probe
is a string worldsheet or a DBI brane action.  The corresponding
Krylov complexity would probe a qualitatively different sector of
the operator space and could display different growth exponents or
transient quiver dynamics. 
Finally, it would be worthwhile to investigate whether similar constructions can be developed for other holographic systems, including theories with reduced supersymmetry or non-conformal dynamics, see for example in backgrounds like \cite{Anabalon:2021tua, Anabalon:2022aig, Anabalon:2024che, Nunez:2023nnl, Nunez:2023xgl, Macpherson:2024qfi, Macpherson:2025pqi, Chatzis:2024top, Chatzis:2025dnu, Chatzis:2025hek, Fatemiabhari:2024aua}.
\section*{Acknowledgments:} Various colleagues have contributed to our understanding and the presentation of the material in this work. We thank: Nicol\'o Bragagnolo (also for collaborations!), Pawel Caputa, Dimitrios Chatzis, Madison Hammond,  Horatiu Nastase, Alfonso Ramallo, Dibakar Roychowdhury, Dimitrios Zoakos.
C. N. is supported by STFC’s grants ST/Y509644-1, ST/X000648/1 and ST/T000813/1. The work of R.T. has been supported by EPSRC Grant EP/Z535175/1 and STFC grant UKRI1787.

\bibliographystyle{JHEP}
\bibliography{main.bib}
\end{document}